\documentclass[prl,twocolumn,,aps,showpacs]{revtex4} 

\usepackage{graphicx}
\usepackage{dcolumn}
\usepackage{bm}

\begin{document}

\title{Statistical mechanics of scale-free networks at a critical point:\\ 
Complexity without irreversibility?}

\author{Christoly Biely$^{1,2}$}
\email[E-mail: ]{christoly.biely@meduniwien.ac.at}
\author{Stefan Thurner$^{1,2}$}
\email[E-mail: ]{thurner@univie.ac.at}
\affiliation{
   $^1$Complex Systems Research Group  HNO; 
          Medical University of Vienna; W\"ahringer G\"urtel 18-20; A-1090 Vienna; Austria \\
   $^{2}$Atominstitut der \"Osterreichischen Universit\"aten; 
          Stadionallee 2; A-1020 Vienna; Austria\\
}

\date{\today}

\begin{abstract}
Based on a rigorous extension of classical statistical mechanics to networks, 
we study a specific microscopic network Hamiltonian.
The form of this Hamiltonian is derived from the assumption that individual 
nodes increase/decrease their utility by linking to nodes with a higher/lower 
degree than their own. We interpret utility 
as an equivalent to energy in physical systems and discuss the temperature 
dependence of the emerging networks. We observe the existence of a critical 
temperature $T_c$ where total energy (utility) and network-architecture 
undergo radical changes. 
Along this topological transition we obtain 
scale-free networks with complex hierarchical topology. 
In contrast to models for scale-free networks introduced so far, 
the scale-free nature emerges within equilibrium, with a clearly defined 
microcanonical ensemble and the principle of detailed
balance strictly fulfilled. This provides clear evidence that 
'complex' networks may arise without irreversibility.
The results presented here should find a wide variety of applications 
in socio-economic statistical systems.
\end{abstract}



\maketitle

Triggered by the vast number of observed non-trivial networks in nature, 
recently a respectable number of models have been introduced 
to understand their statistical properties. Since many  of these networks 
differ considerably from pure random graphs \cite{erdoes_original}, 
the notion of complex networks has emerged which is a well 
established concept nowadays \cite{barabasi02,dorogovtsev03}. 
Perhaps the most apparent property distinguishing such 'complex' 
real-world networks from pure random graphs is their scale-free 
degree distribution $P(k)\sim k^{-\gamma}$, 
which seems to be ubiquitous in nature \cite{dorogovtsev03}. Further, many 
real-world networks exhibit a high amount of clustering, 
and sometimes even a non-trivial dependence of the clustering 
coefficient, $C_i$ of node $i$, when seen as a function of its
degree $k_i$. A power form of $\langle{}C(k)\rangle{}\sim k^{-\delta}$ 
can be associated to the 'complex' topological 
property of  hierarchical clustering \cite{Ravasz}.
Allmost all of the microscopic models proposed to describe such 'complex',  
growing or static,  networks
involve non-equilibrium and evolutionary elements, 
manifesting themselves in different procedures of 
preferential attachment \cite{barabasi_alber,Sneppen,pastor,caldarelli} 
or other structured rewirement schemes \cite{manna,kim,thutsall}.
Further, these procedures often involve the need for non-local information. 
The reasoning behind these approaches has further solidified the notion of 
complex networks. 
The concept of non-equilibrium in the context of networks is so dominant, 
that recently even structured rewirement schemes have entered the
very definition of network-ensembles
\cite{dorogovtsev2,dorogovtsev03}.
Less drastic views of ensembles of networks have 
recently been used to generalize random graphs to networks with arbitrary 
degree \cite{abe}, and to generate scale-free networks by 
appropriately tuning the weights of 'network-Feynman graphs' \cite{burda}.

So far, comparatively little has been done to understand complex networks 
from a purely classical statistical mechanics point of view, 
fully satisfying its foundations
such as equal \emph{a-priori} probabilities. 
Aiming at an explanation of scale-free networks based on microscopic 
interactions, several serious equilibrium approaches 
have been proposed \cite{berg,Newman,vicsek1}. 
In particular topological properties of networks associated with specific
Hamiltonians have been studied \cite{Newman,vicsek1}. 
The Hamiltonians investigated there lead to interesting dynamics, 
but -- to our knowledge -- not to scale-free, complex networks. 

The aim of this paper is to close this fundamental gap by proposing 
a Hamiltonian leading to scale-free, hierarchic networks in thermal
equilibrium. 
The form of the Hamiltonian is derived from 
simple, general, and socio-economically motivated  
assumptions about individual utilities of nodes. 
Nodes act as utility maximizers, such as 
physical particles minimize energy. 
For the greater ease of the exposition of our model
we shall use the notions of utility and inverse energy 
interchangeably.

We consider symmetric networks with a fixed number of distinguishable nodes 
$i=1,...,N$, connected by a fixed number of $\ell=1,...,L$ indistinguishable links.
The network is represented by its adjacency matrix $\bf{c}$, where $c_{ij}=c_{ji}=1$, 
if a link connects nodes $i$ and $j$ and $c_{ij}=c_{ji}=0$, otherwise.
Thus, we define the microcanonical partition function as
\begin{equation}
 \Omega(E,N,L)=\sum_{P({\bf{c}})}\frac{1}{L!}\delta (E-\mathcal{H}({\bf{c}}) )
               \delta (L-\textrm{Tr}{} (\frac{{\bf{c^2}}}{2} ) ) \,\, , 
\label{micro_can}
\end{equation}
with $\mathcal{H}({\bf{c}})$ being the network Hamiltonian and $P({\bf{c}})$ 
denoting all permutations of the $N\times{}N$ adjacency-matrix. 
This definition guarantees that each possible configuration of the 
adjacency matrix is realized with the same \emph{a priori} probability.
The canonical partition function may be obtained by the Laplace transform
of Eq. (\ref{micro_can}), \cite{grandy},  
or via the maximum entropy principle, as shown in \cite{Newman}, 
\begin{equation}
  Z(T,N,L)=\sum_{P({\bf{c}})}\delta \left(L-\frac{Tr({\bf{c}}^2)}{2} \right) 
           e^{-\beta{}\mathcal{H}({\bf{c}})} \quad , 
\label{ens_can}
\end{equation}
using the usual definition of temperature $T\equiv\frac{1}{k \beta}$. 
In simulations the canonical ensemble can be  generated e.g. by  
the Metropolis-algorithm: starting from an adjacency matrix 
${\bf{c}}$, a 'virtual' graph $\hat{\bf{c}}$ is generated by a random
rewirement step. Then, $\hat{\bf{c}}$ is accepted with probability
$p_{\rm accept}=\min(1,\exp[-\beta(\mathcal{H}(\hat{{\bf{c}}})-\mathcal{H}({\bf{c}}))])$. 

\begin{figure}[t]
\begin{center}
 \includegraphics[height=60mm]{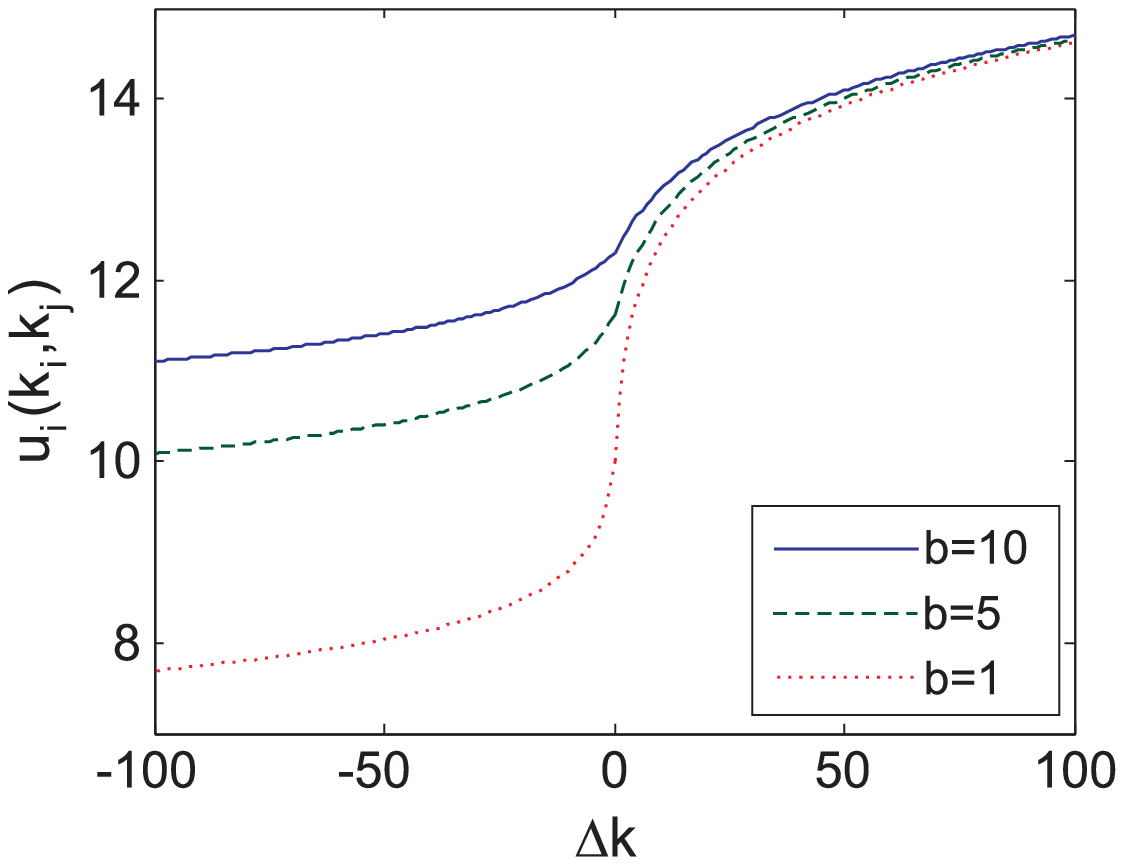}
\end{center}
\caption{Node-utility as a function of $\Delta{}k=k_i-k_j$ for different values of $b$. 
         The parameters in Eq. (\ref{ue_1}) are $a_1=1$, $c_1=10$, 
         $a_2=0.5$ and $c_2=1.5$.}
\label{utility}
\end{figure}


Given Eq. (\ref{ens_can}) one may study any reasonable 
Hamiltonian depending on any properties of the network.
Here, we want to adopt the view of modeling
\emph{microscopic} interactions, 
where the total energy of a network can be expressed as the sum
over all energy/utility contributions of  individual nodes, $u_i$. 
In many realistic settings this node-utility will depend on 
properties/states of node $i$ itself, and on properties of the 
node $j$ whereto a link is established. 
These node-properties are denoted by $\Pi(i)$. 
In this case the Hamiltonian is also expressible
as a sum over all links, 
\vspace{-0.2cm}
\begin{equation}
 \mathcal{H}({\bf{c}})=\sum_{\ell=1}^{L}u_\ell(\Pi(i),\Pi(j))\quad ,
\end{equation}
where $u_\ell(\Pi(i),\Pi(j))$ is the utility of link $\ell$ 
connecting nodes $i$ and $j$, who are characterized by their 
properties $\Pi(i)$ and $\Pi(j)$, respectively.
For simplicity we assume linearity and 
separate $u_{\ell}$ into the individual node-contributions 
$u_\ell(\Pi(i),\Pi(j))=u_i(\Pi(i),\Pi(j))+u_j(\Pi(i),\Pi(j))$. 
In the following we specify the model such that the 
utility of a node increases if it connects to a node 
that is 'more important' than itself. 
Similarly, its utility decreases if it establishes  
a (potentially costly) link to a 'less important' node. 
As a simple measure for importance we suggest the degree 
of a node, i.e. $\Pi(i)=k_i$. The relative importance 
between two nodes is denoted by $\Delta{}k=|k_i-k_j|$, 
which will enter the utility function as the only argument.
For the particular form of the utility function we chose 
a standard, monotonically increasing, concave log-utility 
function \cite{economics}, which incorporates the concept 
of decreasing marginal utility \cite{economics}.
We thus model node-utility by
\begin{eqnarray}
\label{ue_1}
 u_i(k_i,k_j)= \left\{ 
 \begin{array}{ll}
  c_1+a_1\log(b_1+\Delta{k}) &{\textrm{for}\quad k_i>k_j} \\
  c_2-a_2\log(b_2+\Delta{k}) &{\textrm{for}\quad k_i<k_j} 
 \end{array}
 \right.
\end{eqnarray}
with shape parameters $a$ and $b$, and offsets $c$. 
To avoid discontinuity in the utility function
we set $c_2=c_1+a_1\log{}(b_1)+a_2\log{}(b_2)$.
This function is shown in Fig. \ref{utility}.
For the sake of further simplicity, we assume $b_1=b_2=b$,
to obtain a particularly simple form for the link-utility, 
\begin{equation}
  u_{\ell}(k_i,k_j)=c_1+c_2+(a_1-a_2)\log(b+\Delta{}k)\quad .
\label{eq_utility}
\end{equation}
Parameter $c_1$ can be chosen to ensure positive
total utility for each link. Parameter $b$ controlls the 
curvature of the utility function \cite{economics}, and  
will be called  'sensitivity parameter'  in the following.

\begin{figure}[t]
\begin{center}
 \includegraphics[height=60mm]{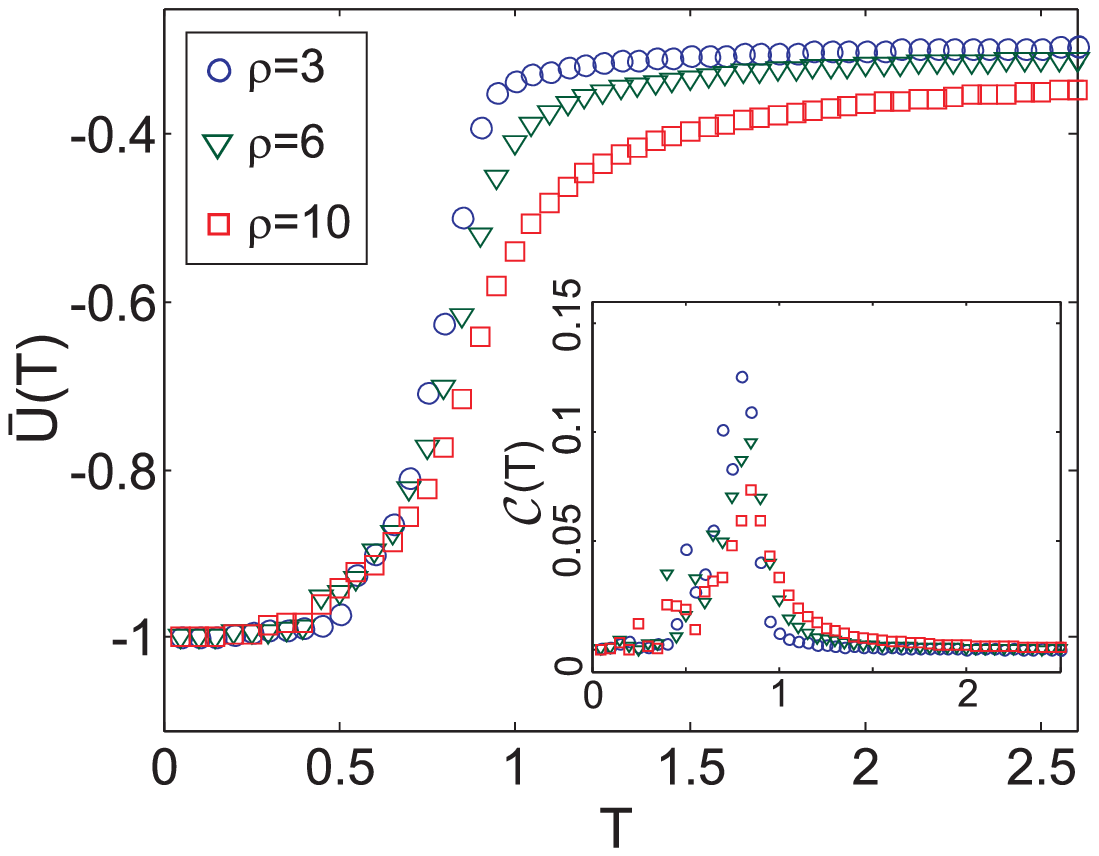}
\end{center}
\caption{Ensemble average of normalized internal energy 
         $\bar{U}(T)=U(T)/\min(U(T))$ and specific heat ${\cal C}(T)$ (inset) as a
         function of $T$, for 
         $N=10^3$, $b=5$ and various densities $\rho$.}
\label{energy}
\end{figure}

Equation (\ref{eq_utility}) can be interpreted as the inverse energy of each 
link which allows us to perform simulations of the  associated  
canonical ensemble, Eq. (\ref{ens_can}). The (collective) 
amount of 'irrationality' of individual nodes, 
i.e. that nodes do not fully maximize their utility (by error or ignorance)  
is captured by the 'temperature' $T$. 
For $a_1=a_2$, the utility is independent of $\Delta k$ 
and we obtain random networks, as expected. For
$a_1\not=a_2$, the constants $a$ and $c$ can be absorbed in
the temperature scale (Boltzmann constant) of the system;  
hence they are omitted without loss of generality.
We assume that $a_1>a_2$,  
meaning that the concave utility-contribution of the node of lower degree
is more dominant than the convex utility-contribution from the node of larger degree,
(putting more emphasis on wins than on losses) 
leading to an asymmetry in utility,  Fig. \ref{utility}.
We finally base our simulations on the Hamiltonian, 
\begin{equation}
 \mathcal{H}({\bf{c}})=-\sum_{\ell}\log(b+\Delta{}k) \quad . 
\label{final_ham}
\end{equation}


\begin{figure}[t]
\begin{center}
 \includegraphics[height=68mm]{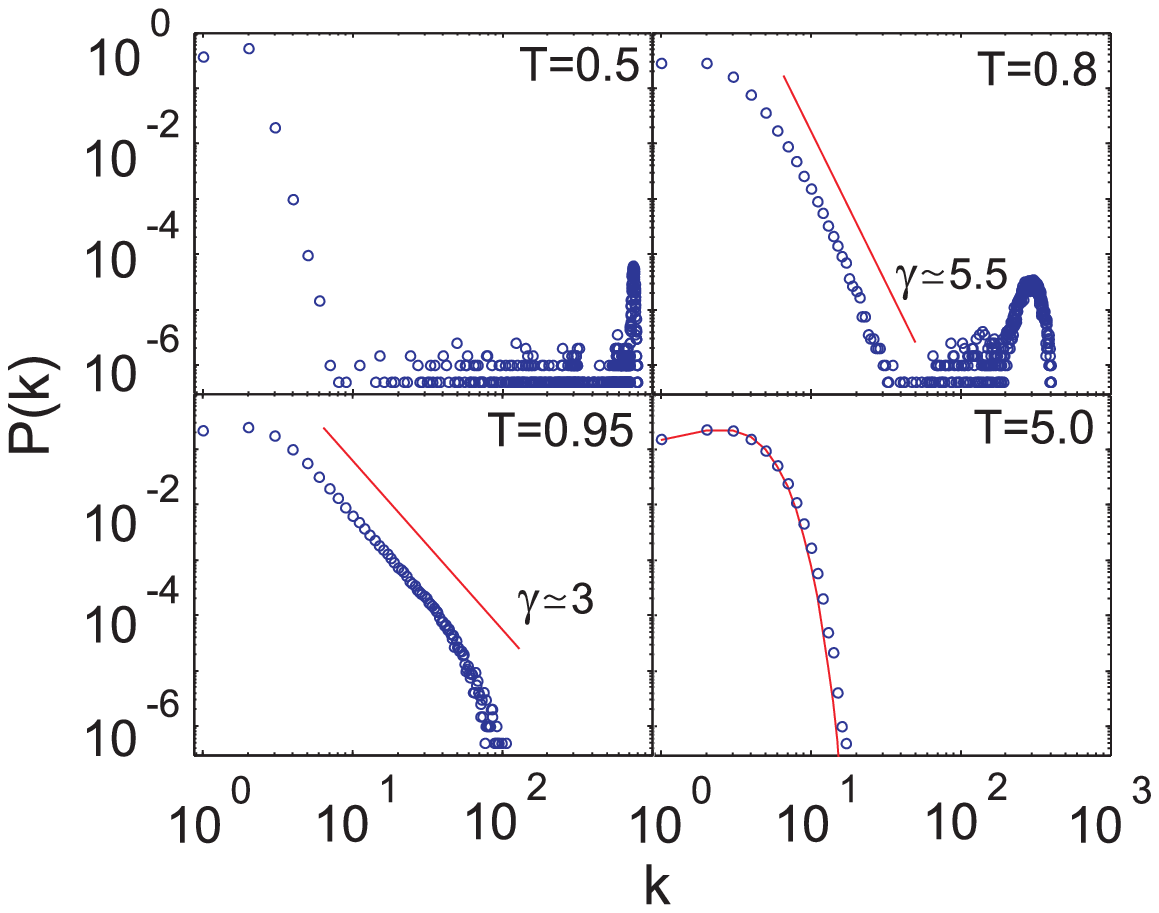}
\end{center}
\caption{Degree distributions at different temperatures for $N=10^3$, 
         $\rho=3$, and $b=5$. 
         The line for $T=5$ is the Poissonian $p(k)=\frac{e^{-\rho} \rho^k}{k!}$. 
} 
\label{degree_T} 
\end{figure}

Based on Eq. (\ref{final_ham}) we simulate networks of the canonical
ensemble, Eq. (\ref{ens_can}), ranging from  $N=500$ to $10^4$ nodes. 
For computational reasons, temperature-dependent results are presented for $N=10^3$. 
All ensemble-averages have been calculated from at least $2\times{}10^3$
configurations, separated by at least $20\times{}N$ update steps.
We analyze the obtained networks as a function of 
the model parameters --  temperature (irrationality) $T$,  link
density $\rho=2L/N$ and  the 'sensitivity' parameter $b$. 

Figure \ref{energy} shows the ensemble average of the total energy of 
the system as a function of $T$ for different values of $\rho$.
For better comparison data has been normalized to the minimum energy.
Also shown is the specific heat ${\cal C}$, 
obtained by a  numeric derivative of the energy-data. 
One clearly finds a radical change in the energy and a characteristic 
maximum of the specific heat at about $T_c=0.8-0.85$,
indicating the presence of a critical point. The transition softens for higher
link-densities, as well as for lower values of $b$ (not shown).
We refrain from commenting on the size of the underlying critical 
exponents, whose proper extraction is beyond the scope of this work.

The change in energy
is associated with considerable restructuring of the underlying networks. To
discuss this in more detail we have calculated ensemble-averages of 
degree distributions for various points along  the transition. 
The results are shown in Fig. \ref{degree_T} for $N=10^3$, $\rho=3$ and
$b=5$. 
For temperatures up to about $T\sim 0.5$, we observe networks with
degrees of all magnitudes. From $T\sim{}0.5$ upward, a core of highly
connected nodes (bump) keeps growing, gradually shifting to the left. 
For $T\lesssim{}0.8$, the two regions in the degree distribution coexist.
In the interval $0.8\lesssim{}T\lesssim{}0.95$, the highly connected core 
merges with the rest of the network and
gradually disappears with further temperature increase.
\begin{figure}[t]
\begin{center}
 \includegraphics[height=60mm]{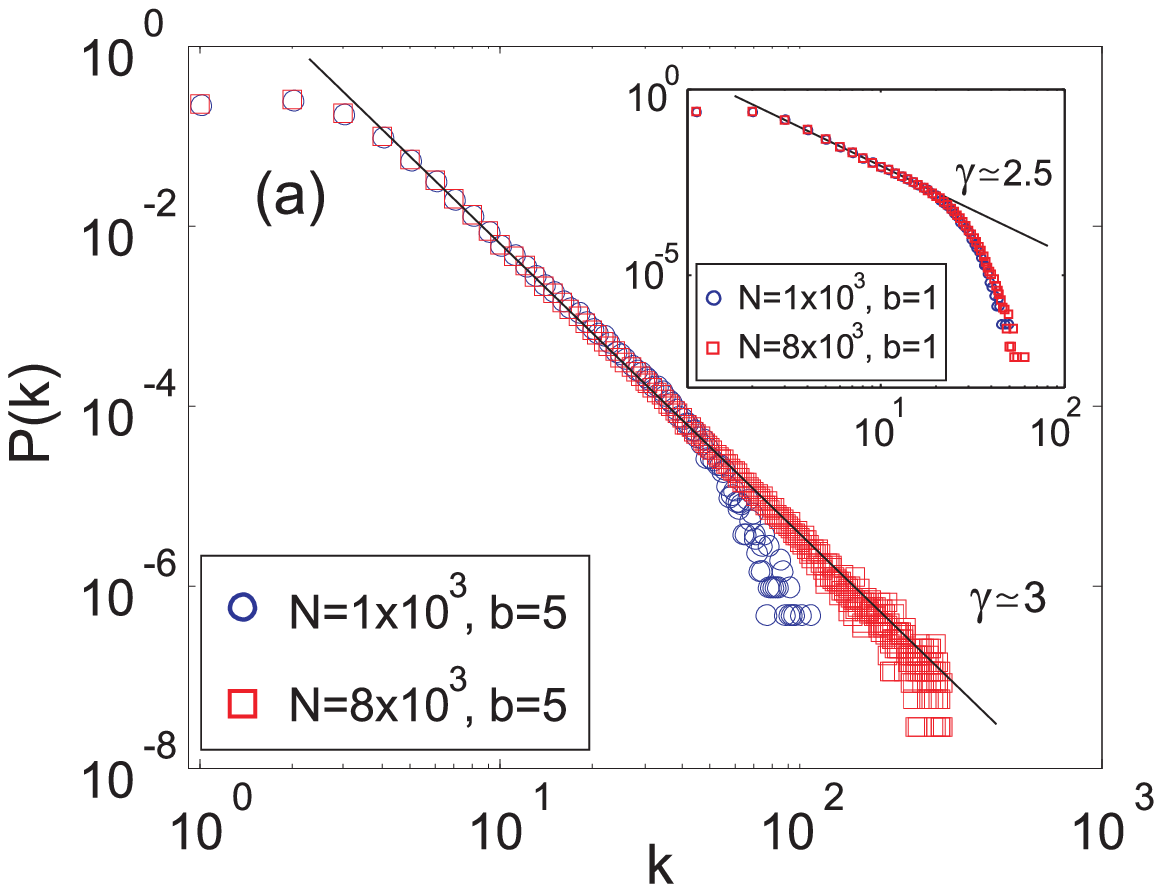}\\
 \hspace{-3mm} \includegraphics[height=62mm]{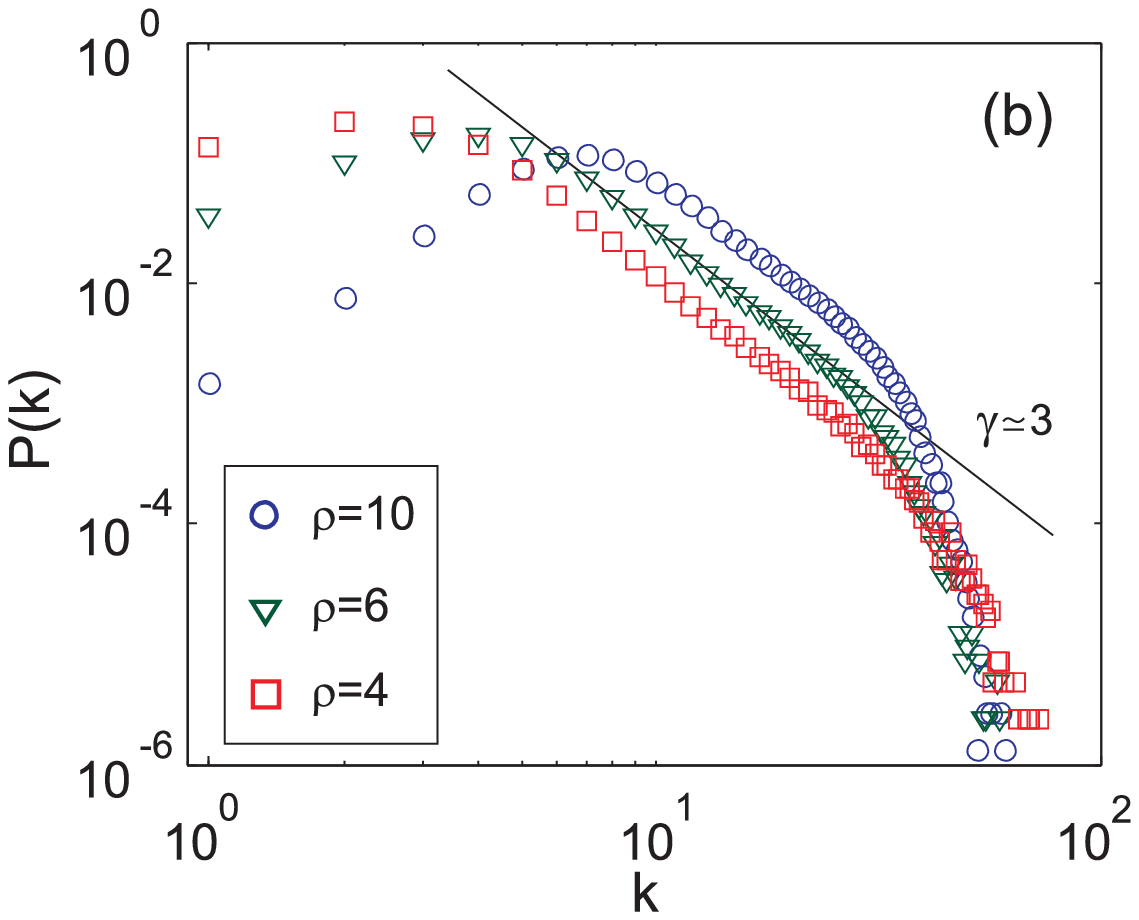}
\end{center}
\caption{(a) Finite size dependence of the degree distribution 
   for various $N$ and $b$  (inset) and $\rho=3$. For $b=5$, we have
   $T=0.95$, being somewhat higher for $b=1$. 
   (b) Degree distribution for different link-densities $\rho$, 
   $N=10^3$, and $b=5$. Temperatures are adjusted to the 
   pure scale-free region.
}
\label{degree_N} 
\end{figure}
At $T\sim0.95$ a pure power-law $P(k)\sim{}k^{-\gamma}$ with exponent $\gamma\sim{}3$ 
matches the degree distribution. 
Ignoring the bump at higher degrees, the degree distribution may also be 
approximated reliably for lower temperatures (down to about $T\sim{}0.80$).
In the $T$ interval $[0.80,0.95]$ the degree exponent covers a range of
$\gamma \in [5.5,3]$, respectively. 
Increasing the temperature above $T=0.95$, keeps the 
power-law exponent $\gamma$ practically unchanged, but shifts the 
exponential cutoff to the left, ultimately leading to random networks
with Poissonian distributions, Fig. \ref{degree_T}. 

Finite-size effects and the role of parameter $b$ for the scale-free region 
are captured in  Fig. \ref{degree_N} a. Sizes $N=10^3$ and $N=8\times{}10^3$
are compared  for $b=1$ and $b=5$; both exhibit nice scaling. 
Power-law fits yield a degree exponent of $\gamma\approx{}3$ and $2.5$
for  $b=5$ and  $b=1$, respectively, regardless of system size.
Variation of $b$ therefor allows to model virtually all exponents 
occurring in  real-world networks \cite{dorogovtsev03}. 

\begin{figure}[t]
\begin{center}
 \includegraphics[height=60mm]{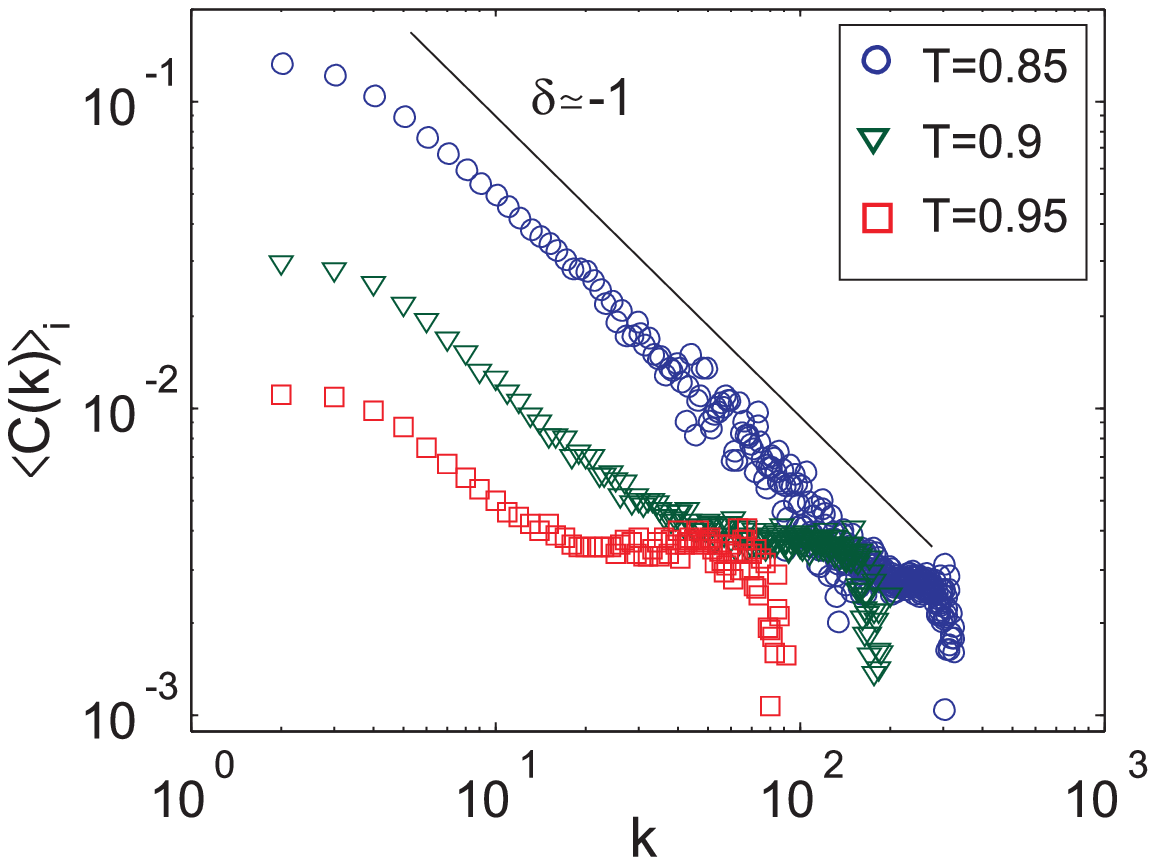}
\end{center}
\caption{Degree-dependence of the average cluster coefficient 
         $\langle{}C(k)\rangle{}$ at different
         temperatures. $N=10^3$, $\rho=3$ and $b=5$. 
} 
\label{pic_clus}
\end{figure}

In Fig. \ref{pic_clus} we show the degree-dependence of ensemble 
averages of the mean cluster coefficient 
$\langle C_i \rangle_i =\langle 2n_i/k_i(k_i-1)\rangle _i$ 
where $n_i$ is the number of links between the 
neighbors of node $i$, 
\footnote{Numerical ensemble averages may
lead to artifacts if $\langle{}C(k)\rangle$, 
in network realizations where the respective 
degree $k$ is not existing, are defined as
$\langle{}C(k)\rangle=0$.}. 
For  $T=0.85$ we obtain a nice scaling-law
$\langle{}C(k)\rangle{}=k^{\delta}$, with $\delta=-1$. 
This is in very good agreement with many empirically
examined data \cite{Ravasz} and demonstrates that our model 
reproduces the 'complex' topological property of hierarchical 
clustering found in many socio-economical networks. 
For higher temperatures, $\delta$ stays the same, however, the 
onset of the cutoff regime changes, resulting in a flat curve 
for high temperatures (not shown). 

The results presented  hold qualitatively for relatively  small $\rho$. 
For $\rho$ larger than $5$, a characteristic scale gradually emerges, 
due to the fact that the mean $\langle{}k\rangle{}$, 
corresponding to high-temperature random networks, 
shifts to larger values. 
Despite this characteristic scale, for appropriate temperatures, a
power-law may still be fitted to a region, Fig. \ref{degree_N} b. 
Here temperatures are chosen such that the pure scale-free region is 
recovered (without the bump). The characteristic exponent of 
$\gamma=3$ is preserved.


In summary, in the course of a very general model of  
socio-economical systems, 
where individuals are utility maximizers with bounded rationality, 
we discovered that scale-free networks 
with hierarchical clustering 
naturally emerge in the vicinity of a critical point.
Most remarkably, the mechanism behind these results is nothing but the
theory of equilibrium statistical mechanics, rigorously applied to networks.
In substantial contrast to work conducted earlier \cite{dorogovtsev2,manna}, 
no modifications of the sampling of phase-space are used
(clearly defined microcanonical ensemble)
such that the full power of equilibrium statistical mechanics is retained. 
We have obtained the first reversible, equilibrium access to 
scale-free networks based on microscopic interactions satisfying detailed balance.
Preferential attachment and structured rewirement schemes 
model many non-equilibrium processes in the real world adequately, 
however, it has to be noted that scale-free networks also exist 
within a pure  equilibrium concept. 
Finally, as the notion of complexity is usually tightly connected to dissipative
structures far from equilibrium, our results could stimulate a 
discussion about the actual complexity of
'complex' networks.

S.T. would like to thank the SFI and in 
particular J.D. Farmer for their great hospitality and support 
in Sept-Oct of 2004. The project was supported by 
the Austrian Science Foundation FWF under P17621 G05.


\vspace{-0.2cm}

\begin{thebibliography}{99}



\bibitem{erdoes_original}
P. Erd\"os, A. R\'enyi, 
Publ. Math. Debrecen {\bf 6}, 290 (1959);
Publ. Math. Inst. Hung. Acad. Sci.  {\bf 5}, 17 (1960).

\bibitem{barabasi02}
A.-L. Barabasi,
Rev.  Mod. Phys. {\bf 74}, 47 (2002).

\bibitem{dorogovtsev03}
S. Dorogovtsev, J.F.F. Mendes,
{\it Evolution of Networks: From Biological Nets to the Internet and WWW},
(Oxford University Press, 2003).

\bibitem{Ravasz}
E. Ravasz, A.-L. Barabasi,
Phys. Rev. E {\bf 67}, 026112 (2003).

\bibitem{barabasi_alber}
A.-L. Barabasi, R. Albert,
Science {\bf 286}, 509 (1999).

\bibitem{Sneppen}
M. Rosvall, K. Sneppen, 
Phys. Rev. Lett. {\bf 91}, 178701 (2003).

\bibitem{pastor}
J.J. Ramasco, S.N. Dorogovtsev and R. Pastor-Satorras,
Phys. Rev. E {\bf 70}, 036106 (2004).

\bibitem{caldarelli}
G. Caldarelli, A. Capocci, P. De Los Rios and M.A. Munoz,
Phys. Rev. Lett. {\bf 89}, 25 (2002).

\bibitem{manna}
M. Baiesi, S.S. Manna,
Phys. Rev. E {\bf 68}, 047103 (2003).

\bibitem{kim}
B.J. Kim, A. Trusina, P. Minnhagen and K. Sneppen, 
Eur. Phy. J. B {\bf 43}, 369 (2005).

\bibitem{thutsall}
S. Thurner, C. Tsallis, 
cond-mat/0506140. 

\bibitem{dorogovtsev2}
S.N. Dorogovtsev, J.F.F. Mendes and A.N. Samukhin,
Nucl. Phys. B {\bf 666}, 396 (2003).

\bibitem{abe}
S. Abe, S. Thurner, 
cond-mat/0501429.

\bibitem{burda}
Z. Burda, J.D. Correia and A. Krzywicki,
Phys. Rev. E {\bf 64}, 046118 (2001).

\bibitem{berg}
J. Berg, M. L{\"a}ssig,
Phys. Rev. Lett. {\bf 89}, 228701 (2002).

\bibitem{Newman}
J. Park, M.E.J. Newman,
Phys. Rev. E {\bf 70}, 066117 (2004).

\bibitem{vicsek1}
I. Farkas, I. Derenyi, G. Palla and T. Vicsek,
Lect. Notes Phys. {\bf 650}, 163 (2004);
G. Palla, I. Derenyi, I. Farkas and T. Vicsek,
Phys. Rev. E {\bf 69}, 046117 (2004).

\bibitem{grandy}
W.T. Grandy, 
{\it Foundations of Statistical Mechanics},
(Kluwer Academic Publishers, 1987).

\bibitem{economics}
J.E. Ingersoll,
{\it Theory of Financial Decision Making},
(Rowman \& Littlefield Publishers, 1987).


\end{thebibliography}
\end{document}